\documentclass[runningheads]{llncs}
\usepackage[T1]{fontenc}
\usepackage{graphicx}
\usepackage{float}
\usepackage{subcaption} 
\usepackage{multirow,booktabs,array,makecell}
\usepackage{tikz}
\usepackage{hyperref}
\usepackage{placeins}
\usepackage{appendix}
\usepackage[T1]{fontenc}
\usepackage[htt]{hyphenat}
\usetikzlibrary{positioning}
\newcolumntype{L}[1]{>{\raggedright\arraybackslash}p{#1}}
\newcolumntype{L}[1]{>{\raggedright\arraybackslash}p{#1}}
\begin{document}
\title{Engineering Spatial and Molecular Features from Cellular Niches to Inform Predictions of Inflammatory Bowel Disease}
\titlerunning{Engineered Cellular Niche Features Predict IBD}
%
\author{Myles Joshua Toledo Tan\inst{1,2}\orcidID{0000-0002-1426-6526} \and
Maria Kapetanaki\inst{3}
\and
Panayiotis V. Benos\inst{2}\orcidID{0000-0003-3172-3132}} 
\authorrunning{M.J.T. Tan et al.}
%
\institute{Department of Electrical and Computer Engineering, Herbert Wertheim College of Engineering, University of Florida, Gainesville, Florida, 32611, United States
\email{mylesjoshua.tan@medicine.ufl.edu}
\and
Department of Epidemiology, College of Public Health and Health Professions and College of Medicine, University of Florida, Gainesville, Florida, 32610, United States
\email{pbenos@ufl.edu}
\and
Department of Pharmacotherapy and Translational Research, College of Pharmacy, University of Florida, Gainesville, Florida, 32603, United States of America\\
\email{mkapetanaki@ufl.edu}}
\maketitle              
\begin{abstract}
Differentiating between the two main subtypes of Inflammatory Bowel Disease (IBD): Crohn’s disease (CD) and ulcerative colitis (UC) is a persistent clinical challenge due to overlapping presentations. This study introduces a novel computational framework that employs spatial transcriptomics (ST) to create an explainable machine learning model for IBD classification. We analyzed ST data from the colonic mucosa of healthy controls (HC), UC, and CD subjects. Using Non-negative Matrix Factorization (NMF), we first identified four recurring cellular niches, representing distinct functional microenvironments within the tissue. From these niches, we systematically engineered 44 features capturing three key aspects of tissue pathology: niche composition, neighborhood enrichment, and niche-gene signals. A multilayer perceptron (MLP) classifier trained on these features achieved an accuracy of $0.774 \pm 0.161$ for the more challenging three-class problem (HC, UC, and CD) and $0.916 \pm 0.118$ in the two-class problem of distinguishing IBD from healthy tissue. Crucially, model explainability analysis revealed that disruptions in the spatial organization of niches were the strongest predictors of general inflammation, while the classification between UC and CD relied on specific niche-gene expression signatures. This work provides a robust, proof-of-concept pipeline that transforms descriptive spatial data into an accurate and explainable predictive tool, offering not only a potential new diagnostic paradigm but also deeper insights into the distinct biological mechanisms that drive IBD subtypes.

\end{abstract}
\section{Introduction}
Inflammatory Bowel Disease (IBD) presents a significant clinical challenge due to the diagnostic and therapeutic ambiguity between its two main subtypes, Crohn's disease (CD) and ulcerative colitis (UC) \cite{IBDdefinition1,IBDdefinition2}. Though pathologically distinct, overlapping clinical presentations can complicate diagnosis, which is critical as treatment strategies diverge substantially \cite{UC}. Subject heterogeneity in presentation and therapeutic response underscores the need for precise diagnostic tools reflecting underlying cellular and molecular complexity \cite{UCthera}.

Single-cell RNA sequencing (scRNA-seq) has provided unprecedented insight into this complexity, revealing diverse immune, stromal, and epithelial cell states in the inflamed gut \cite{scrna1,scrna2,scrna3}. Garrido-Trigo et al. combined scRNA-seq with spatial imaging to highlight that the greatest inter-patient variability in IBD lies within myeloid cells, particularly macrophages and neutrophils \cite{IBDpaper}. While essential for unraveling disease mechanisms at a cell-by-cell resolution \cite{Gudino}, dissociation-based scRNA-seq loses native tissue architecture. Spatial transcriptomics (ST) preserves this, enabling gene expression analysis in its morphological context \cite{SpTxStahl,SpTxRao}. Here, we use ST to investigate whether the organization of cells into functional microenvironments can serve as a robust diagnostic tool for IBD subtypes.

We hypothesized that disruptions in the structure of \textit{cellular niches}\footnote{In this article, the term \textit{niche} is used to mean a latent factor. This is a data-derived unit representing recurring patterns of cellular organization inferred from NMF, rather than a predefined biological microenvironment.}, which are localized communities of interacting cells, are distinct hallmarks of UC and CD. Our computational approach classified cell types \cite{cell2location} to map cell populations across ST data from colonic tissue, then applied non-negative matrix factorization (NMF) to identify four recurring cellular niches defined by unique cell type combinations. This analytical framework, which deconstructs spatial data into functional units, has proven effective in identifying pathological microenvironments in other complex inflammatory conditions like idiopathic pulmonary fibrosis \cite{MayrIPF}. Similar ST approaches have mapped healing programs in mouse models of colitis \cite{Parigi}, fibrosis-associated networks in stricturing CD \cite{Kong}, and cellular ecosystems correlating with UC therapeutic response \cite{Mennillo}. To build a predictive model, we engineered 44 features capturing three key aspects of the tissue: niche composition (relative abundance), neighborhood enrichment (spatial interactions), and niche-gene signals (localized gene expression). A multilayer perceptron (MLP) classifier was then trained on these features to distinguish between healthy controls (HC), UC, and CD.


Our work builds on the foundational spatial characterizations of IBD \cite{IBDpaper} and niche decomposition frameworks in fibrosis \cite{MayrIPF} by introducing a multilayered feature engineering strategy. Although recent landmark studies have focused on characterizing spatial landscapes \cite{Parigi,Kong} or correlating them with treatment outcomes \cite{Mennillo}, our approach represents a critical next step: transforming these descriptive spatial insights into a robust predictive model capable of distinguishing IBD subtypes. This approach not only provides a potential new diagnostic paradigm, but also offers deeper insights into the distinct biological mechanisms that drive UC and CD, linking cellular atlases to the context of functional tissue and paving the way for more targeted therapeutic strategies. 

Given the methodological complexity of this article, we provide an overview of the computational pipeline in Figure~\ref{fig:IBD_flowchart} (Appendix).

\section{Methods}
\subsection{Data}

The study used two publicly available datasets from the NCBI Gene Expression Omnibus (GEO). The primary dataset, GSE234713 \cite{IBDpaper,CosMx_data}, consisted of ST data from colonic mucosa. This data was collected from nine formalin-fixed paraffin-embedded (FFPE) human samples, including three healthy non-IBD controls (HC), three subjects with CD, and three with UC. Generated using the NanoString CosMx \cite{NanostringCosMx} platform, this dataset provided expression profiles for 980 genes. The secondary dataset, GSE214695 \cite{IBDpaper,scRNA_data}, contained scRNA-seq data from a separate cohort of healthy and IBD colonic mucosa samples. These scRNA-seq data served as a reference for classifying and assigning cell types within the primary spatial dataset. A detailed breakdown of the dataset, including the conditions (health or disease states), FFPE samples per state, number of fields of view (FOV) per sample, and number of cells per sample, is provided in Table~\ref{Data}.

\begin{table}[!h]
\centering
\caption{Overview of IBD CosMx NanoString data from colonic mucosa}\label{Data}
\begin{tabular}{|l|l|r|r|}
\hline
\textbf{Condition Groups} & \textbf{FFPE Samples} & \textbf{\# of FOVs} & \textbf{\# of Cells} \\
\hline
healthy controls (HC) & HC a & 19 & 39,101 \\
 & HC b & 20 & 54,059 \\
 & HC c & 16 & 27,905 \\
 & \textit{\textbf{HC total}} & \textit{\textbf{55}} & \textit{\textbf{121,065}} \\
\hline
ulcerative colitis (UC) & UC a & 19 & 49,240 \\
 & UC b & 22 & 76,613 \\
 & UC c & 21 & 54,811 \\
 & \textit{\textbf{UC total}} & \textit{\textbf{62}} & \textit{\textbf{180,664}} \\
\hline
Crohn’s disease (CD) & CD a & 19 & 31,582 \\
 & CD b & 19 & 72,440 \\
 & CD c & 16 & 53,344 \\
 & \textit{\textbf{CD total}} & \textit{\textbf{54}} & \textit{\textbf{157,366}} \\
\hline
\textbf{3 Condition Groups} & \textbf{9 FFPE Samples} & \textbf{171 FOVs} & \textbf{459,095 Cells} \\
\hline
\end{tabular}
\end{table}

\subsection{Cell type classification}

Cell type classification was performed using \texttt{cell2location} \cite{cell2location} to infer the probability of individual cell types across the ST dataset. The CosMx dataset and a scRNA-seq reference dataset were first aligned by identifying common genes between the two datasets ($n = 976$ shared genes). Genes with low expression were removed using the following thresholds: a minimum of five cells expressing the gene, a non-zero mean expression $\geq1.12$ counts, and 3\% minimum expression frequency. Mitochondrial genes were excluded to reduce noise.  

The \texttt{cell2location} \cite{cell2location} regression model was trained on the reference scRNA-seq dataset to estimate a gene expression signature matrix $S_{g \times c}$, where $g=871$ is the number of genes and $c=54$ is the number of cell types. Training was performed with 500 epochs, a learning rate of 0.002, and a batch size of 2,500 cells.  

The trained model was then applied to the spatial data to solve for $W_{s \times c}$, representing the probability of each cell type $c$ at each spatial location $s$, by maximizing a Gamma-Poisson likelihood. This produced high-resolution spatial maps of cell type distributions for downstream analysis.  

\subsection{Cellular niche decomposition}

Cellular niches were identified by applying NMF \cite{NMF} to the inferred cell type probability matrix $W_{s \times c}$. The goal was to decompose this matrix into two low-rank, non-negative matrices:  

\[
W_{s \times c} \approx U_{s \times k} \cdot H_{k \times c}
\]

where $k$ is the number of latent factors (niches), $U$ represents the contribution of each spatial location to a niche, and $H$ represents the relative composition of cell types within each niche.  

The number of factors was set to $k = 4$, based on the elbow method applied to the reconstruction error curve. The factorization was optimized using the non-negative double singular value decomposition (NNDSVD) \cite{NNDSVD} initialization with a maximum of 1,000 iterations and a random seed of 0 for reproducibility.  

Each cell was assigned to its dominant niche by $\arg\max(U_{s,k})$. This process reveals the cellular composition of each of the four niches (latent factors) and allows for the assignment of each individual cell to its dominant niche, providing a basis for subsequent feature engineering. 

\subsection{Feature engineering}

A set of 44 features that was used for downstream classification consisted of three groups: (i) four niche composition features representing the relative abundance of each niche within a FOV, (ii) 16 neighborhood enrichment features capturing the spatial relationships between niches, and (iii) 24 niche--gene features identified through an information-theoretic selection process. Together, these features comprehensively represent the biological, spatial, and molecular characteristics of each FOV and served as input to the MLP classifier described in a later section.

\subsubsection{Niche composition.} 
Each cell was assigned to one of four NMF-derived niches. 
For each field-of-view (FOV), the niche composition features were computed as the proportion of cells in each niche:
\[
\mathrm{comp\_niche}_i = \frac{n_i}{\sum_{k=1}^4 n_k},
\]
where \(n_i\) is the number of cells in Niche \(i \in \{1,2,3,4\}\). 
This produced four features per FOV.

\subsubsection{Niche neighborhood enrichment score}

\begin{figure}[!h]
    \centering
    \includegraphics[width=0.3\textwidth]{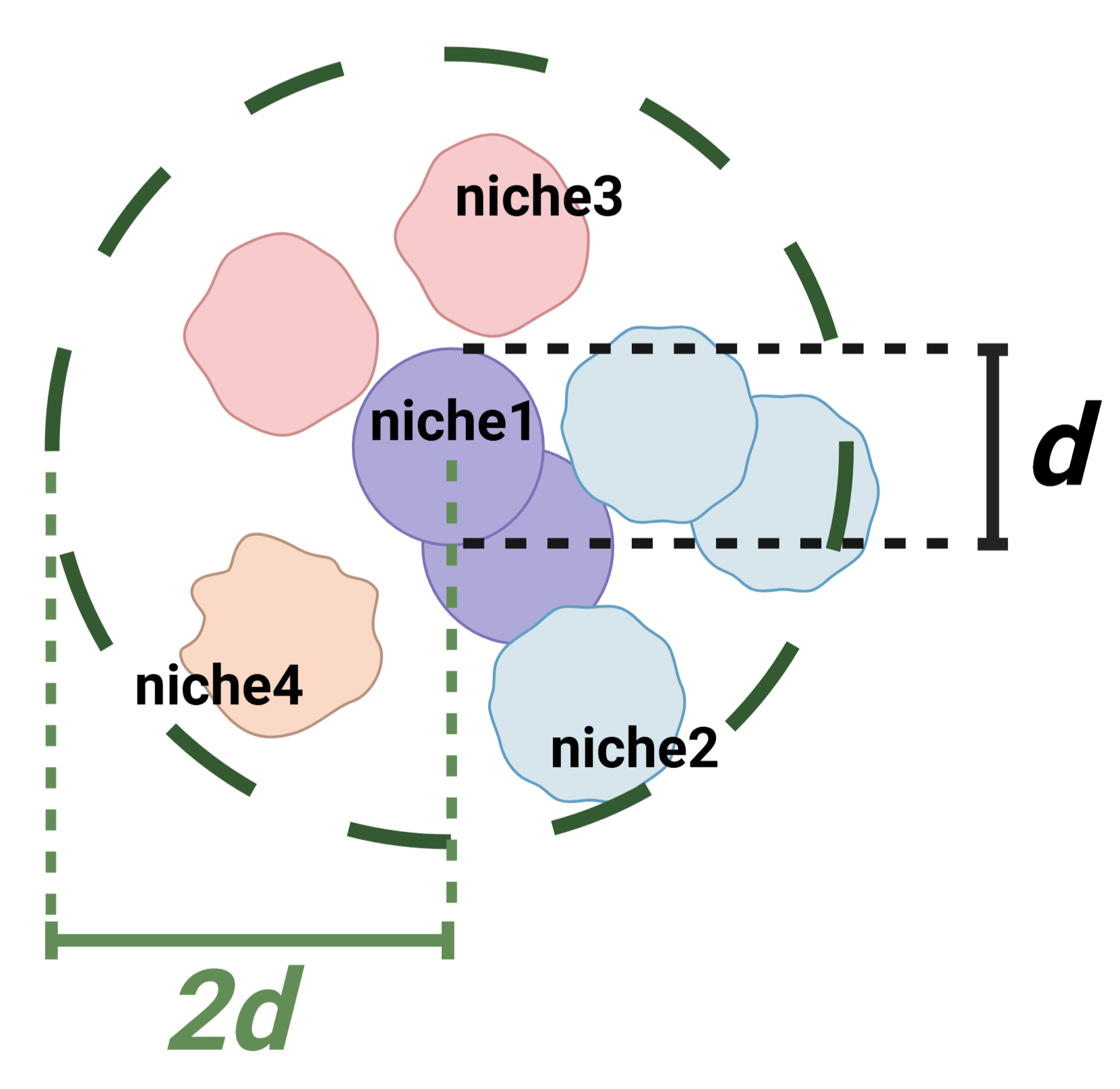}
    \caption{Illustration of cell--cell neighborhood centered on a focal cell (Niche 1, purple). 
    The focal cell has diameter \textit{d}, so any cell within a radius $2d$ from the center of the focal cell is part of its neighborhood.
    In this example, the neighborhood includes one additional Niche 1 cell, three Niche 2 cells (blue), two Niche 3 cells (pink), and one Niche 4 cell (orange). 
    The dashed green circle represents the neighborhood boundary. 
    Created in \href{https://BioRender.com}{https://BioRender.com}.}
    \label{fig:cell_neighborhood}
\end{figure}

To capture local spatial organization, we examined cell--cell neighborhoods within each FOV. 
A KD-tree \cite{KDTree} was used to identify neighbors, where any cell within a distance equal to twice the diameter of the focal cell was considered a neighbor (Figure~\ref{fig:cell_neighborhood}). 
Let \(O_{i,j}\) be the observed count of ordered neighbor pairs in which the focal cell belongs to Niche \(i\) and the neighboring cell to Niche \(j\) (\(i,j \in \{1,2,3,4\}\)). 
Let \(p_i\) represent the overall proportion of Niche \(i\) cells in the FOV and let \(T = \sum_{i=1}^4 \sum_{j=1}^4 O_{i,j}\) be the total number of ordered pairs.
The expected count under random mixing is \(E_{i,j} = T \cdot p_i p_j\).
A Laplace-smoothed log-ratio was then calculated to quantify enrichment $\mathrm{S}_{i,j}$:
\[
\mathrm{S}_{i,j} = \log_2\left(\frac{O_{i,j} + 1}{E_{i,j} + 1}\right).
\]
This generated a total of 16 features per FOV (a \(4 \times 4\) matrix including self-pairs).

\subsubsection{Niche--gene features with information-theoretic selection}

For each niche, gene-level signals were aggregated within every FOV by taking the mean expression of each gene across all cells assigned to that niche, producing features labeled as 
\texttt{niche\_[niche\#]\_gene\_[gene]}. 
To identify the most informative subset of these features for distinguishing among the three condition groups (HC, UC, and CD) we used mutual information (MI) \cite{MutualInfo} between each feature \(X\) and a binary class label \(Y_d\) (1 for the condition group \(d\), 0 otherwise).

For discrete variables, the MI is:
\[
I(X;Y_d) = \sum_{x} \sum_{y \in \{0,1\}} p(x, y) \log_2 \left(\frac{p(x, y)}{p(x)\,p(y)}\right),
\]
where \(p(x, y)\) is the joint probability of feature value \(x\) and label \(y\), while \(p(x)\) and \(p(y)\) are the corresponding marginal probabilities.

The top 15 features with the highest MI were selected for each group (45 total). 
For each selected feature, a two-sided Mann--Whitney U test \cite{MW1,MW2} was then used to compare its distribution between FOVs labeled as \(d\) and those labeled as non-\(d\). 
Multiple testing correction was performed using the Benjamini--Hochberg false discovery rate (FDR) \cite{FDR}, and any features appearing in more than one disease list were removed to ensure disease specificity. 
This process yielded 24 unique, statistically significant niche--gene features.

\subsection{Multilayer perceptron}

A MLP \cite{MLP1,MLP2} was trained to classify each field of view (FOV) into one of three condition groups: HC, UC, or CD. The model input consisted of 44 engineered features (\(p = 44\)), and the output was a three-class categorical variable (\(k = 3\)). A pipeline was constructed to standardize features followed by MLP classification. Hyperparameter optimization \cite{hyperparam} was performed using randomized search over 30,000 candidate configurations with a three-fold stratified group cross-validation scheme (\(n = 171\) FOVs) with subject ID as the grouping variable, i.e., all FOVs from a given subject were assigned to the same fold. The search explored activation functions (\(\mathrm{ReLU}\) \cite{relu} or \(\tanh\) \cite{tanh}), regularization strengths \(\alpha \sim \mathrm{LogUniform}(10^{-5}, 10^{-1})\), batch sizes \(\{2, 4, 8, 16\}\), and seven hidden layer architectures: \((25)\) [single small layer], \((32, 16, 8)\) [progressively tapering], \((40, 20, 10, 5)\) [deeper with small neurons], \((44, 22)\) [starting at feature count, then tapering], \((50, 25, 12)\) [moderately wide tapering], \((50, 50)\) [uniform width], and \((64, 32)\) [wide, shallow network].

\begin{figure}[htbp]
    \centering
    \includegraphics[width=0.95\textwidth]{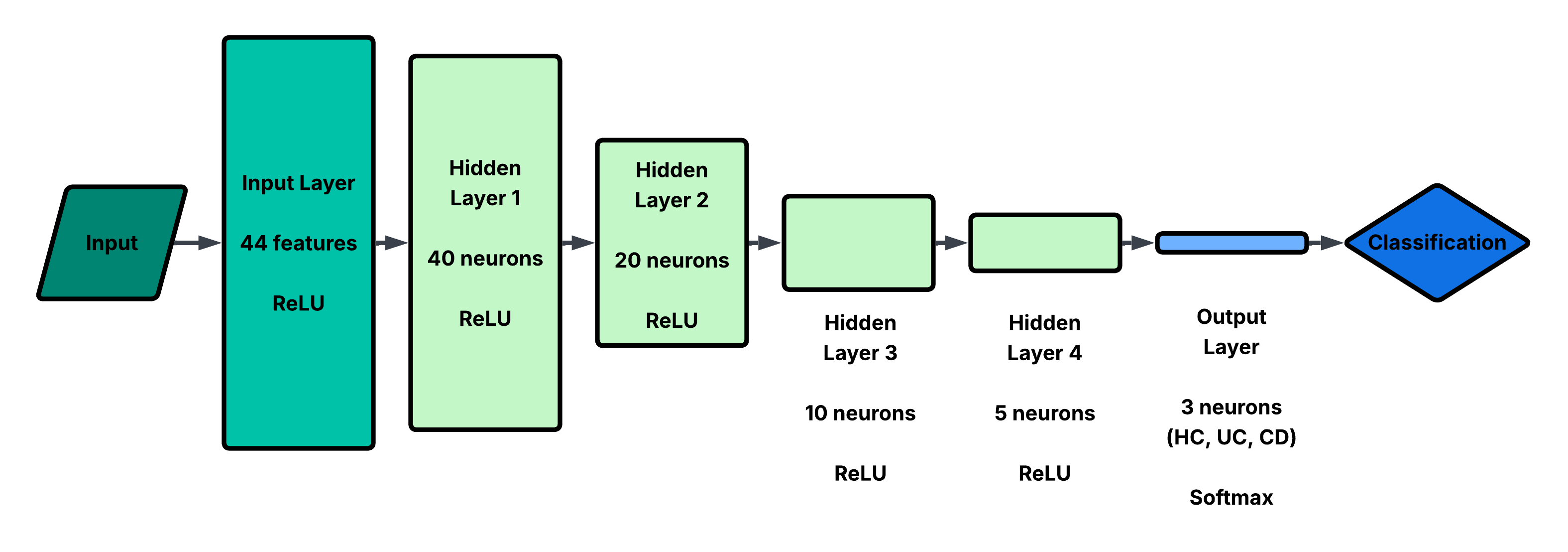}
    \caption{Architecture of the multilayer perceptron (MLP) used for classification. 
    The network consists of an input layer with 44 features, four hidden layers 
    with 40, 20, 10, and 5 neurons respectively, and an output layer with three neurons 
    (corresponding to HC, UC, and CD) using the softmax activation function.}
    \label{fig:mlp_architecture}
\end{figure}

The optimal architecture (Figure~\ref{fig:mlp_architecture}) consisted of four hidden layers with the number of neurons decreasing in size: \(40 \rightarrow 20 \rightarrow 10 \rightarrow 5\). It used the ReLU activation function, a batch size of 4, and an \(L^2\) regularization \cite{L2} parameter $\alpha = 0.001$ (rounded; results not sensitive to finer precision). This configuration achieved a mean F1-score of 0.712 for the three-class problem. The model was optimized with Adam \cite{adam} using \texttt{MLPClassifier} in scikit-learn with \texttt{learning\_rate='adaptive'}, which reduces the base learning rate when training plateaus. This is the scheduler implemented by \texttt{MLPClassifier} and is distinct from the per-parameter adaptation inherent to Adam \cite{adam}. A maximum of 1000 iterations was used. Performance was evaluated using weighted F1-score as the primary metric, with additional reporting of accuracy, precision, and recall \cite{metrics}. Final evaluation was based on mean performance across folds and included a confusion matrix to assess misclassification patterns across condition groups.

\subsection{Model explainability analysis}
\subsubsection{Causal discovery}

Potential cause-effect relationships among features and condition group (Disease/Health State) were inferred using the Fast Causal Inference (FCI)-Stable algorithm \cite{FCI1,FCI2} implemented in the \texttt{rCausalMGM} R package \cite{MGM1,MGM2}. Three separate datasets were analyzed: (i) niche composition (four variables representing the log-transformed proportions of each niche per field of view), (ii) 16 neighborhood enrichment features, and (iii) 24 niche--gene features identified through the information-theoretic approach. Each dataset included the categorical condition group variable HC, UC, CD) as a target node.  

For each dataset, FCI-Stable was run with a significance level of $\alpha = 0.05$ and the orientation rule set to \texttt{maxp} \cite{maxp}. This algorithm identifies a partial ancestral graph (PAG), where nodes represent either spatial niches, enrichment variables, or niche--gene features, and edges encode potential direct or conditional dependencies. Analyses were performed separately for each condition group and for the combined disease variable, yielding graphical models that capture directional relationships between features and disease. The resulting causal graphs were exported as SIF files for downstream visualization and interpretation.

\subsubsection{Statistical tests}

To assess whether the composition of niches differed across the three groups, the percentage of cells belonging to each NMF factor was calculated for every field of view (FOV). Because the data did not follow a normal distribution, non-parametric statistical methods were employed. Pairwise group comparisons were performed using Dunn’s post-hoc test to evaluate differences between HC, UC, and CD for each niche factor. The Benjamini--Hochberg FDR \cite{FDR} correction was applied to account for multiple testing. For each comparison, the difference in group means was computed, and the direction of change was indicated as ``Up'' if \(\mathrm{Group 1} > \mathrm{Group 2}\) or ``Down'' otherwise. Comparisons with adjusted \(p\)-values \(<0.05\) were considered statistically significant. 

Neighborhood enrichment was quantified for each FOV by comparing observed versus expected counts of adjacent niche interactions. These enrichment scores were then analyzed to determine whether the spatial organization of niches differed across HC, UC, and CD. A global Kruskal--Wallis test \cite{KW} was first performed for each niche interaction to evaluate overall differences among the three condition groups. When a global test reached significance (\(p < 0.05\)), pairwise Mann--Whitney U tests \cite{MW1,MW2} were subsequently conducted to identify specific group-level differences. The Benjamini--Hochberg procedure \cite{FDR} was again used to correct for multiple comparisons. Significant pairwise comparisons were reported with their adjusted \(p\)-values, and the relative direction of change (``Up'' or ``Down'') indicated whether the enrichment of a given niche interaction was higher or lower in one group compared to another.

\subsubsection{Feature importance analysis}
To interpret the MLP classifier, permutation importance (PI) \cite{FeatImp1,FeatImp2} was computed to quantify the contribution of each feature to model performance. For each feature, its values were randomly permuted across samples while keeping other features constant. The resulting decrease in weighted F1-score was recorded. This procedure was repeated multiple times to estimate the mean and standard deviation (SD) of the F1-score change.

The absolute value of the mean decrease was used to rank features, with the sign of the original mean indicating whether the feature had a positive (blue) or negative (red) association with correct classification. Error bars represented the variability (SD) across permutations.

This analysis was performed for both the three-class task (HC, UC, CD) and a two-class task where UC and CD were combined into a single IBD class. The three-class results identified features distinguishing all three conditions, while the binary classification highlighted features most critical for separating HC from IBD. Together, these analyses provided complementary insights into the most influential biological and spatial predictors among the 44 input features.

\section{Results}
\subsection{Cell type classification and cellular niche decomposition}

\begin{figure}[htbp]
    \centering
    \begin{subfigure}[b]{0.45\textwidth}
        \centering
        \includegraphics[width=\textwidth]{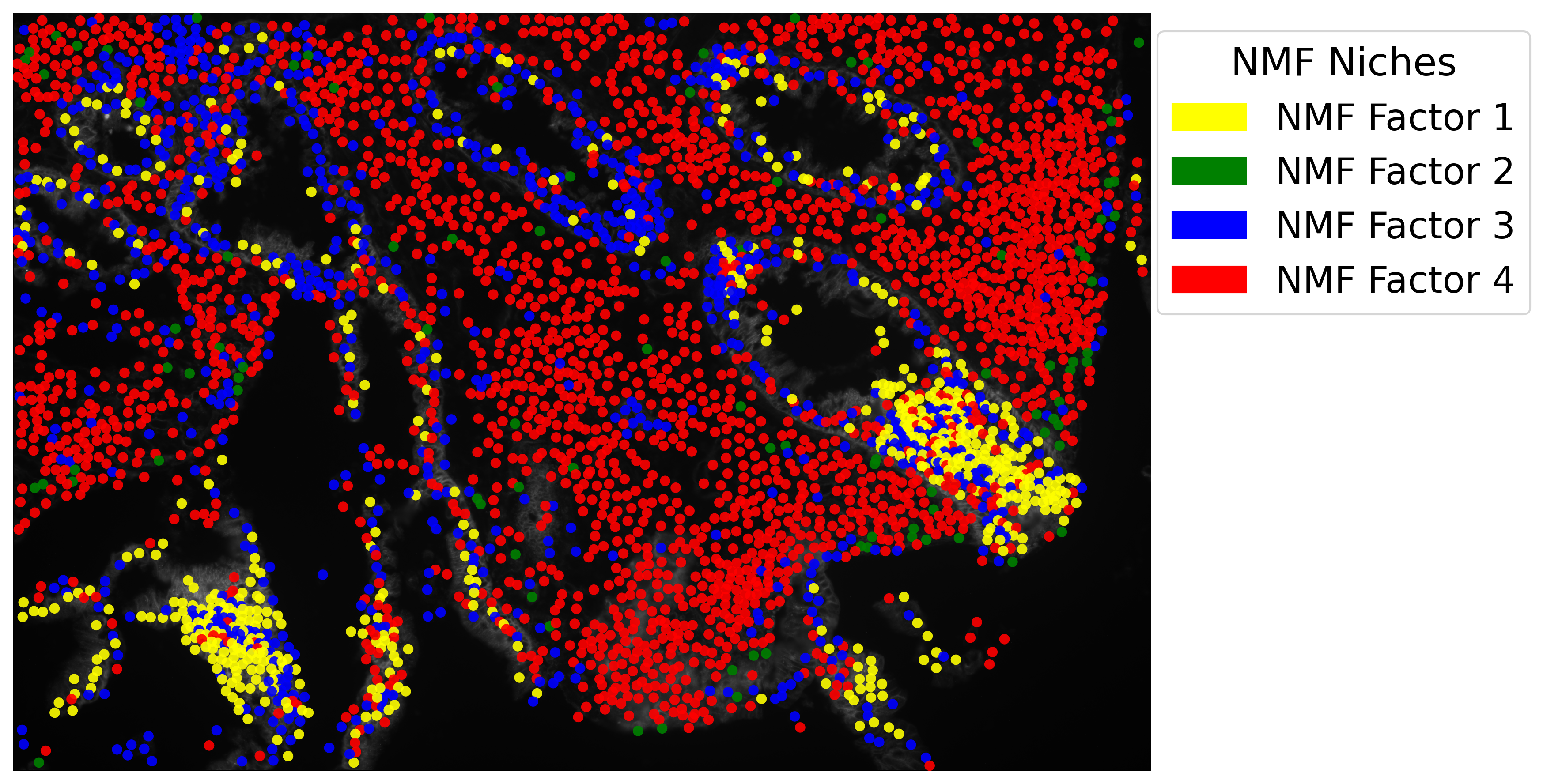}
        \caption{}
        \label{fig:nmf_mapping}
    \end{subfigure}
    \hfill
    \begin{subfigure}[b]{0.51\textwidth}
        \centering
        \includegraphics[width=\textwidth]{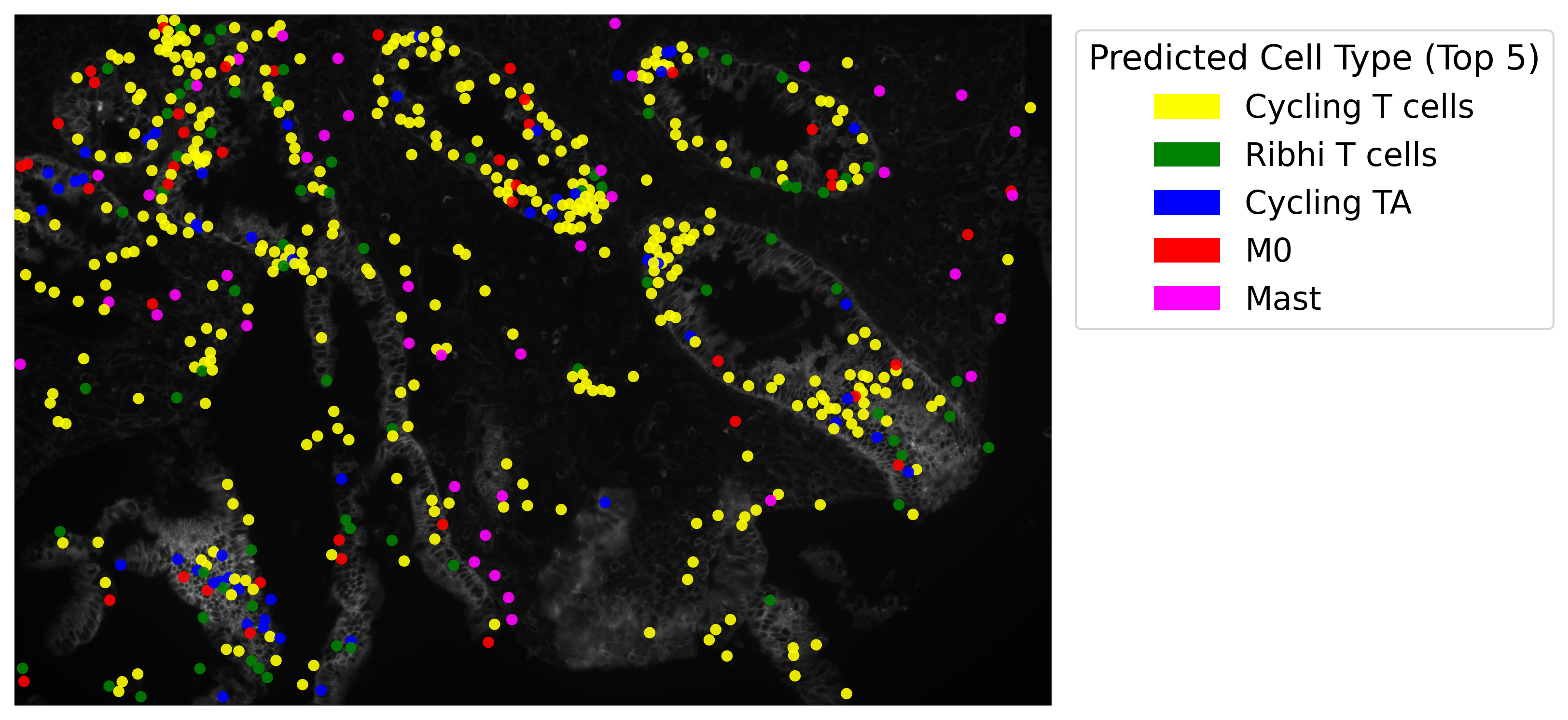}
        \caption{}
        \label{fig:top5_celltypes}
    \end{subfigure}

    \caption{
        Visualization of cellular niches and their cell type composition in FOV \texttt{UC a\_8}. 
        (a) Cellular niches identified by NMF as colored points overlaid on the histomorphology; 
        (b) The five most abundant cell types within Niche 3.
    }
    \label{fig:nmf_and_celltypes}
\end{figure}

Figure~\ref{fig:nmf_and_celltypes} illustrates how our pipeline links computational niche discovery with biological interpretation using one representative field of view (FOV), \texttt{UC a\_8}.  

In panel (a), each point represents a single cell overlaid on the raw histomorphology image. Colors indicate NMF factors, which define distinct cellular niches. This visualization shows how niches are spatially arranged within the tissue and how they relate to the underlying morphology. The color legend allows straightforward identification of niche boundaries and neighboring regions.

Panel (b) focuses on Niche 3 identified in panel (a). Here, cells are colored by their predicted cell type, with only the five most abundant cell types shown. Each color represents a unique cell type, as indicated in the legend, revealing how these populations are distributed within the niche.  

Together, these panels demonstrate how computational results can be anchored in biological context. Panel (a) shows the spatial layout of inferred niches, while panel (b) links one niche to its cellular composition. Although only one FOV is presented, this approach can be applied across larger datasets to uncover biologically meaningful spatial patterns and guide downstream interpretation.

\subsection{Feature engineering}
\subsubsection{Niche neighborhood enrichment score}

\begin{figure}[htb]
    \centering
    \begin{subfigure}[t]{0.32\textwidth}
        \centering
        \includegraphics[width=\textwidth]{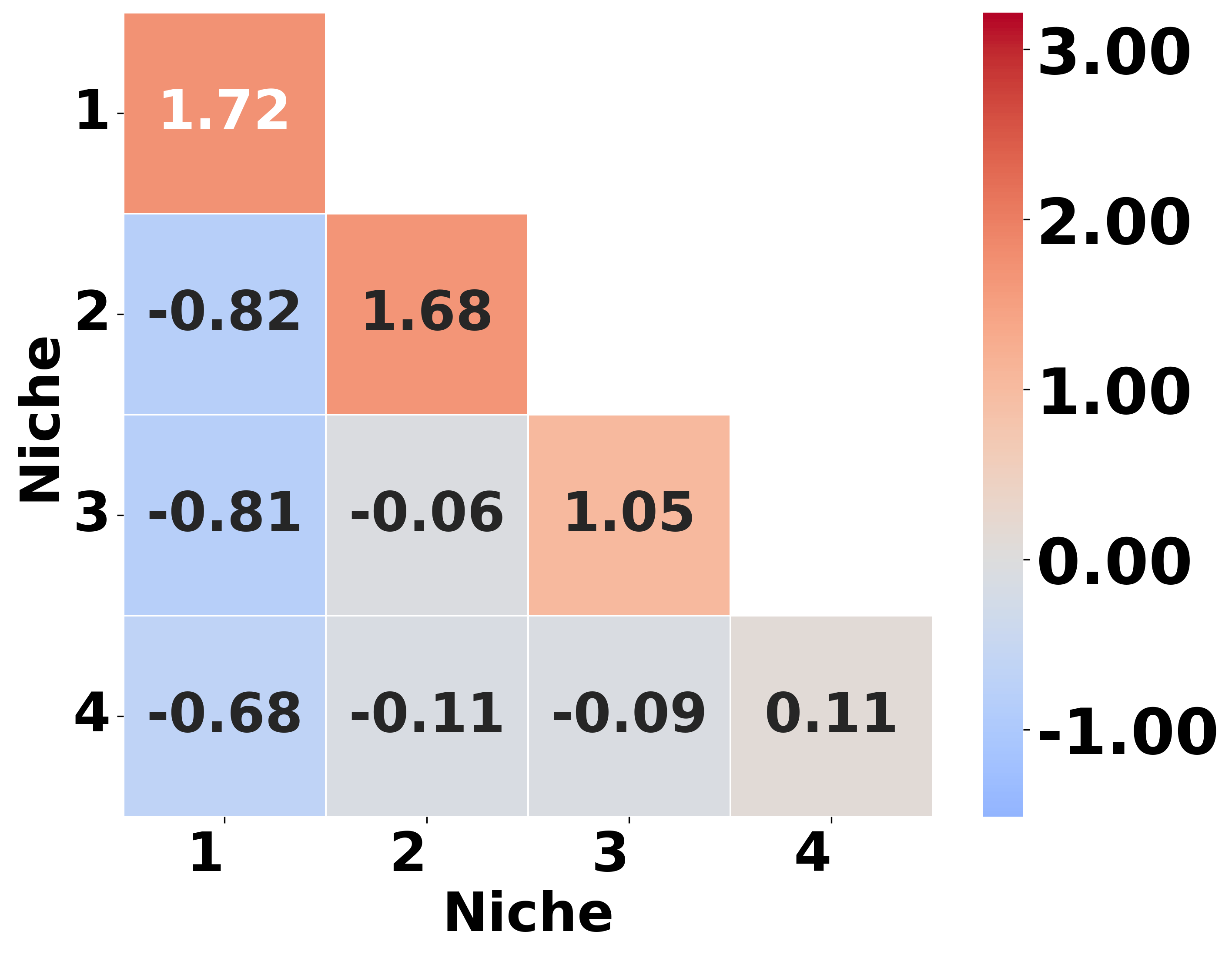}
        \caption{}
        \label{fig:hc}
    \end{subfigure}
    \hfill
    \begin{subfigure}[t]{0.32\textwidth}
        \centering
        \includegraphics[width=\textwidth]{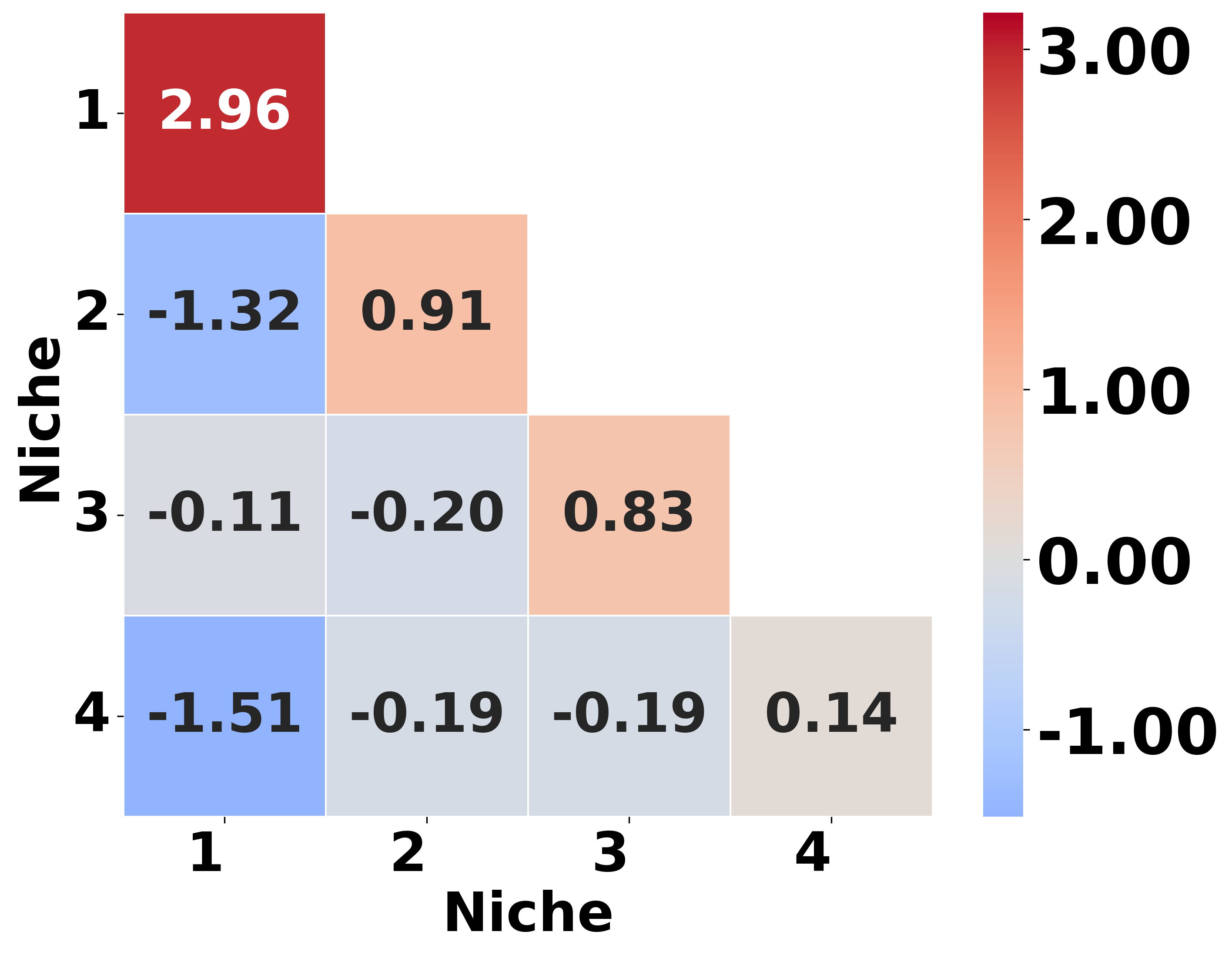}
        \caption{}
        \label{fig:uc}
    \end{subfigure}
    \hfill
    \begin{subfigure}[t]{0.32\textwidth}
        \centering
        \includegraphics[width=\textwidth]{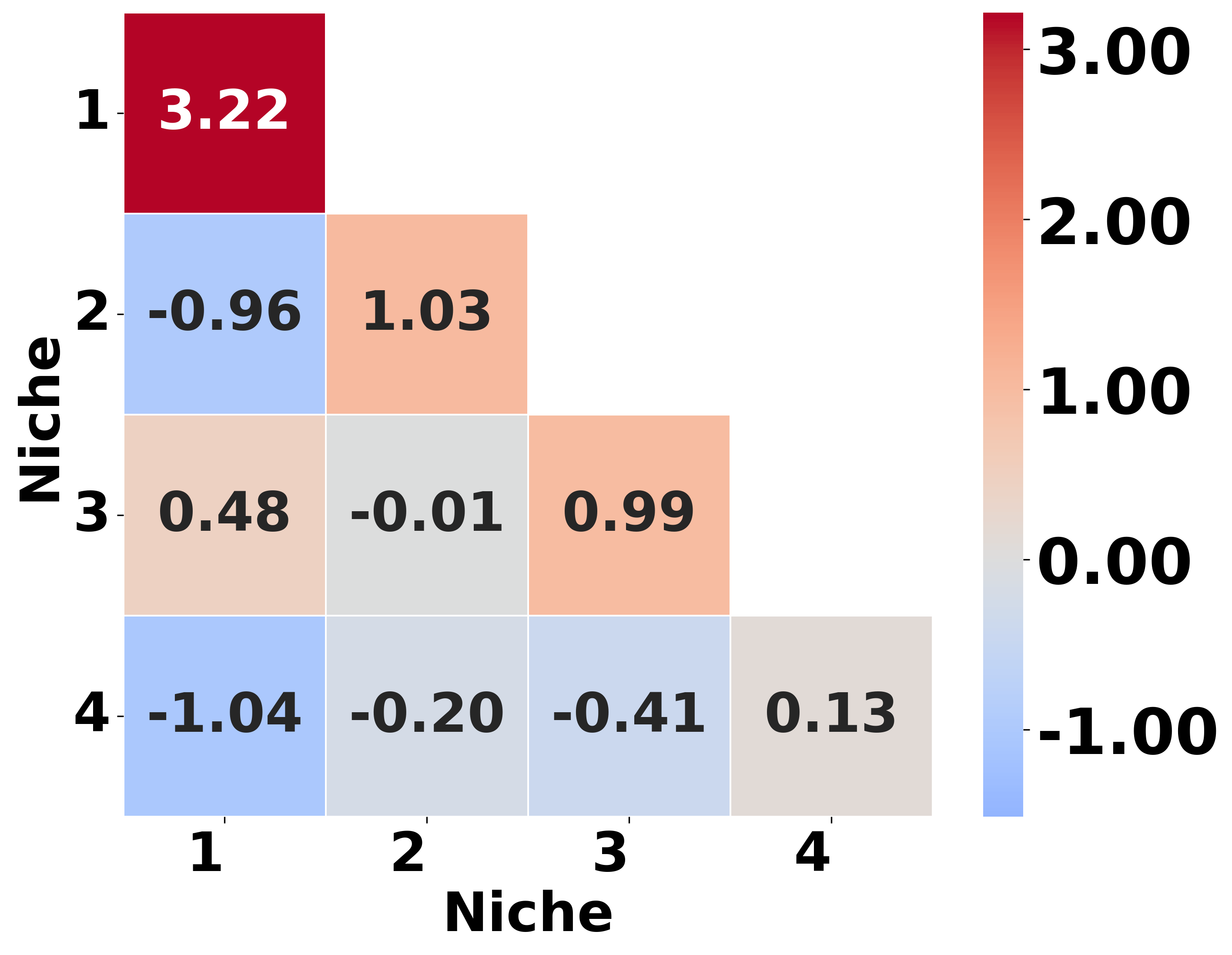}
        \caption{}
        \label{fig:cd}
    \end{subfigure}

    \caption{Niche enrichment comparisons across 
    (a) HC, 
    (b) UC, 
    (c) CD.}
    \label{fig:niche_enrichment}
\end{figure}

Figure~\ref{fig:niche_enrichment} shows niche neighborhood enrichment scores across the three condition groups. Each heatmap compares observed versus expected interactions between pairs of niches within a given FOV. Red indicates niche pairs that co-occur more frequently than expected, while blue indicates niche pairs that are less frequently observed. The diagonal elements reflect interactions within the same niche, while off-diagonal elements capture relationships between different niches.  

This analysis highlights how the spatial organization of cellular niches varies across condition groups. For example, Niche Pair \texttt{1,3} is enriched in CD but not in HC and healthy tissue, suggesting disease-specific alterations in tissue structure. These enrichment patterns provide insight into how disease progression reshapes the spatial context of the cellular microenvironment.

\subsubsection{Niche--gene features with information-theoretic selection}

A total of 24 niche-gene features were identified (Table \ref{tab:mi_scores}). HC-associated features were related to the genes \textit{PIGR}, \textit{IGHG2}, \textit{HLA-DRB1}, \textit{IGHG1}, \textit{CD38}, \textit{STAT1}, and \textit{CHI3L1}
expressed in Niche 1, 2, and 4 cells; while UC-associated features were related to the genes \textit{IGKC}, \textit{IGFBP5}, \textit{HLA-DQB1}, \textit{CASP3}, and \textit{COL3A1} expressed in Niche 2, 3, and 4 cells; and CD-associated features were related to the genes \textit{MZT2A}, \textit{COL9A2}, \textit{FZD1}, \textit{HDAC5}, \textit{COL5A1}, \textit{SOX6}, \textit{FOXF1} expressed in Niche 3 and 4 cells. These features were integrated with niche composition and neighborhood enrichment features to form the final 44-feature dataset used for classification. Table \ref{tab:mi_scores} also indicates whether each gene has been previously reported in the literature to be associated with IBD. When marked “yes,” the corresponding reference cites the specific study supporting that association.

\begin{table}[!h]
\centering
\caption{Significant niche-gene features selected through MI score analysis}
\label{tab:mi_scores}
\begin{tabular}{lcccc}
\hline
\textbf{Niche-Gene Feature} &
\shortstack{\textbf{MI}\\\textbf{Score}} &
\shortstack{\textbf{Condition}\\\textbf{Group}} &
\shortstack{\textbf{Adj.}\\\textbf{\textit{p}-value}} &
\shortstack{\textbf{Known IBD}\\\textbf{Association}} \\ 
\hline
niche\_1\_gene\_HLA-DRB1   & 0.351 & HC & 0.037 & yes~\cite{HLA-DRB1} \\
niche\_1\_gene\_IGHG2      & 0.351 & HC & <0.001 & yes~\cite{IGHG}\\
niche\_1\_gene\_PIGR       & 0.357 & HC & <0.001 & yes~\cite{PIGR}\\
niche\_2\_gene\_IGFBP5     & 0.271 & UC & <0.001 & yes~\cite{IGFBP5}\\
niche\_2\_gene\_IGHG1      & 0.348 & HC & <0.001 & yes~\cite{IGHG}\\
niche\_2\_gene\_IGHG2      & 0.352 & HC & <0.001 & yes~\cite{IGHG}\\
niche\_3\_gene\_COL9A2     & 0.230 & CD & 0.020 & no\\
niche\_3\_gene\_HLA-DQB1   & 0.261 & UC & <0.001 & yes~\cite{HLA-DQB1}\\
niche\_3\_gene\_IGFBP5     & 0.276 & UC & <0.001 & yes~\cite{IGFBP5}\\
niche\_3\_gene\_MZT2A     & 0.244 & CD & <0.001 & no\\
niche\_4\_gene\_CASP3      & 0.280 & UC & <0.001 & yes~\cite{CASP3}\\
niche\_4\_gene\_CD38       & 0.401 & HC & <0.001 & yes~\cite{CD38}\\
niche\_4\_gene\_CHI3L1     & 0.363 & HC & <0.001 & yes~\cite{CHI3L1}\\
niche\_4\_gene\_COL3A1     & 0.267 & UC & <0.001 & yes~\cite{COL3A1}\\
niche\_4\_gene\_COL5A1     & 0.252 & CD & <0.001 & yes~\cite{COL5A1}\\
niche\_4\_gene\_FOXF1      & 0.211 & CD & 0.011 & no\\
niche\_4\_gene\_FZD1       & 0.352 & CD & <0.001 & yes~\cite{FZD1}\\
niche\_4\_gene\_HDAC5      & 0.262 & CD & 0.004 & yes~\cite{HDAC5}\\
niche\_4\_gene\_HLA-DQB1   & 0.277 & UC & <0.001 & yes~\cite{HLA-DQB1}\\
niche\_4\_gene\_IGHG2      & 0.369 & HC & <0.001 & yes~\cite{IGHG}\\
niche\_4\_gene\_IGKC       & 0.313 & UC & 0.001 & yes~\cite{IGKC}\\
niche\_4\_gene\_MZT2A      & 0.209 & CD & 0.002 & no\\
niche\_4\_gene\_SOX6       & 0.235 & CD & 0.010 & yes~\cite{SOX6}\\
niche\_4\_gene\_STAT1      & 0.367 & HC & <0.001 & yes~\cite{STAT1}\\
\hline
\end{tabular}
\end{table}

\subsection{Multilayer perceptron}

\begin{figure}[htb]
\centering
\begin{subfigure}[b]{0.38\textwidth}
    \centering
    \includegraphics[width=\textwidth]{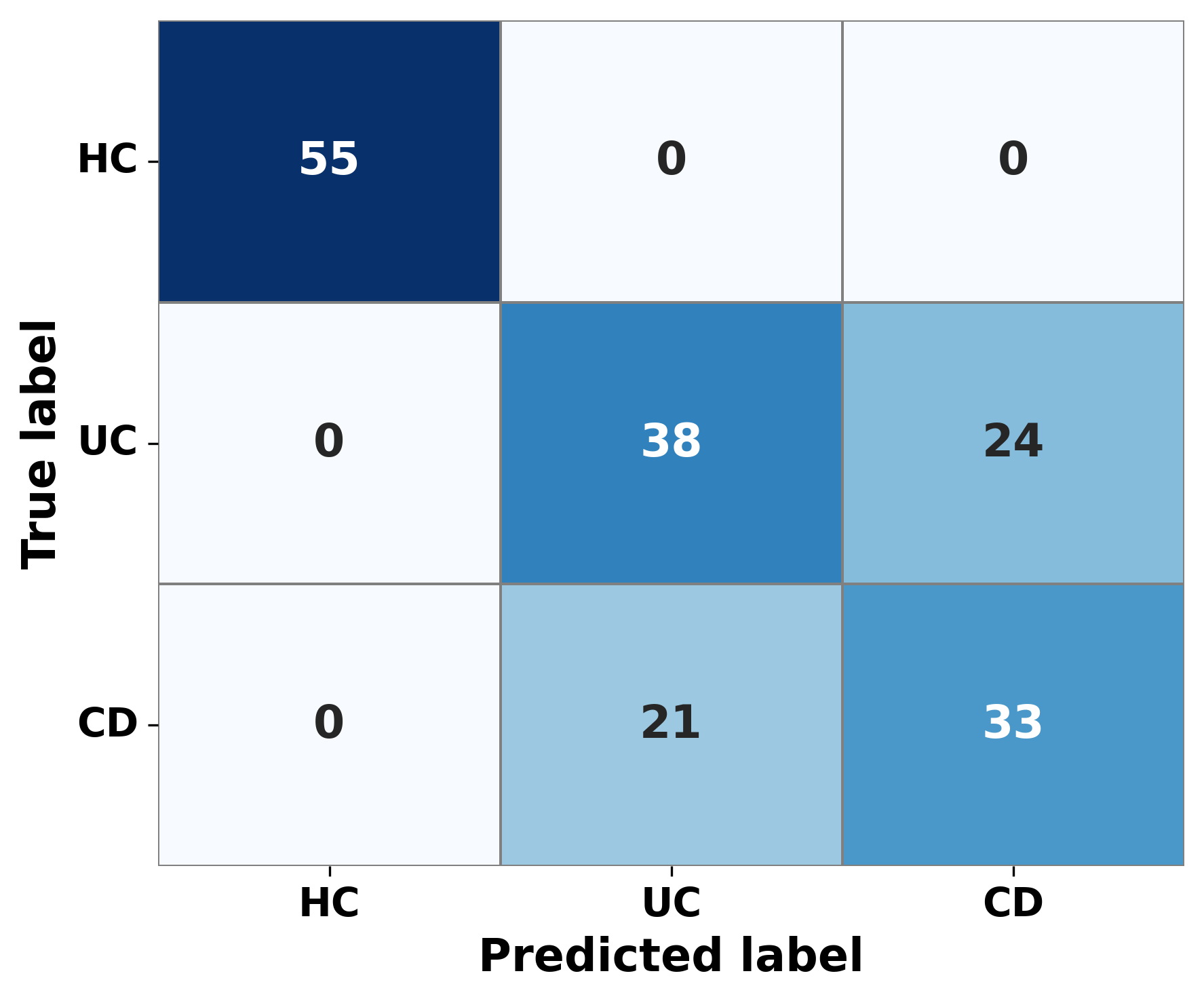}
    \caption{}
    \label{fig:cm_threeclass}
\end{subfigure}
\hfill
\begin{subfigure}[b]{0.32\textwidth}
    \centering
    \includegraphics[width=\textwidth]{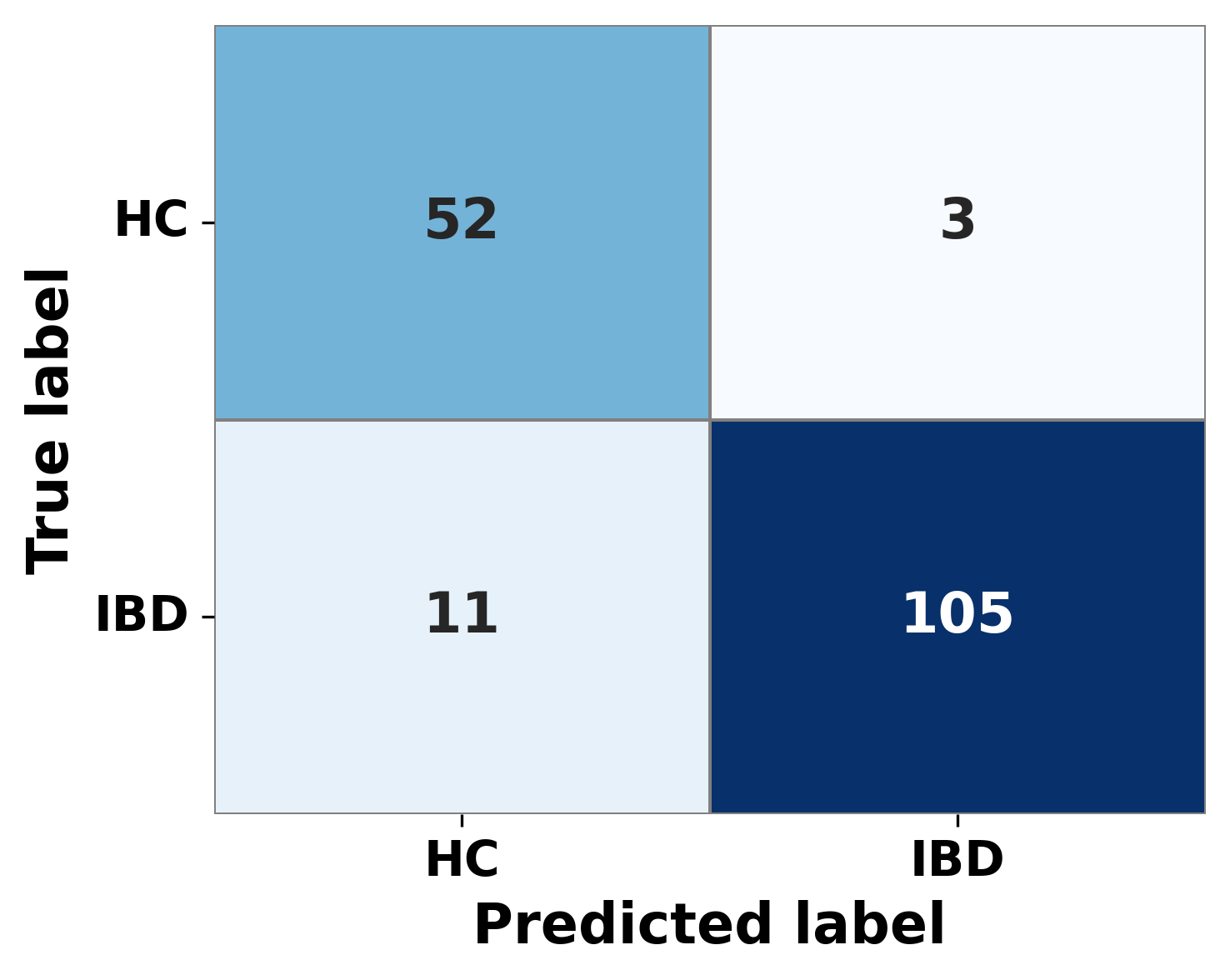}
    \caption{}
    \label{fig:cm_binary}
\end{subfigure}
\caption{Confusion matrices for the (a) three-class, and (b) two-class problems.}
\label{fig:confusion_matrices}
\end{figure}

Figure~\ref{fig:confusion_matrices} shows the confusion matrices for both classification tasks, obtained by aggregating predictions across all three folds of group-stratified cross-validation (subjects were held out by group in each fold). In the three-class case (Figure~\ref{fig:cm_threeclass}), HC samples were classified perfectly, while misclassifications primarily occurred between UC and CD, indicating overlapping features between these two disease states. In the two-class case (Figure~\ref{fig:cm_binary}), the model showed strong separation between HC and IBD, with only a small number of misclassifications.

The corresponding classification reports, computed from the aggregated cross-validation results, are provided in Tables~\ref{tab:classification_report_3class} and~\ref{tab:classification_report_2class}. Table~\ref{tab:performance} summarizes the mean and standard deviation of each performance metric across the three cross-validation folds, providing a measure of variability in model performance. For the more challenging three-class problem, the macro-averaged F1-score was 0.741 (accuracy $0.774\pm0.161$), with perfect HC classification and errors between UC and CD, reflecting their biological similarity. In comparison, the two-class problem showed excellent separation of HC and IBD, with overall accuracy of $0.916\pm0.118$.

Overall, these results indicate that while distinguishing between UC and CD remains challenging due to biological similarity, the MLP classifier is highly effective at differentiating IBD subjects from HC.

\begin{table}[!h]
\centering
\caption{Classification report for the three-class problem (HC, UC, CD).}
\label{tab:classification_report_3class}
\begin{tabular}{lcccc}
\hline
\textbf{Class} & \textbf{Precision} & \textbf{Recall} & \textbf{F1-Score} & \textbf{Support} \\
\hline
HC & 1.000 & 1.000 & 1.000 & 55 \\
UC & 0.644 & 0.613 & 0.628 & 62 \\
CD & 0.579 & 0.611 & 0.595 & 54 \\
\hline
\textbf{Accuracy}     &       &       & 0.737 & 171 \\
\textbf{Macro Avg}    & 0.741 & 0.741 & 0.741 & 171 \\
\textbf{Weighted Avg} & 0.738 & 0.737 & 0.737 & 171 \\
\hline
\end{tabular}
\end{table}

\begin{table}[!h]
\centering
\caption{Classification report for the two-class problem (HC, IBD)}
\label{tab:classification_report_2class}
\begin{tabular}{lcccc}
\hline
\textbf{Class} & \textbf{Precision} & \textbf{Recall} & \textbf{F1-Score} & \textbf{Support} \\
\hline
HC   & 0.825 & 0.945 & 0.881 & 55 \\
IBD  & 0.972 & 0.905 & 0.938 & 116 \\
\hline
\textbf{Accuracy}     &       &       & 0.918 & 171 \\
\textbf{Macro Avg}    & 0.899 & 0.925 & 0.909 & 171 \\
\textbf{Weighted Avg} & 0.925 & 0.918 & 0.919 & 171 \\
\hline
\end{tabular}
\end{table}

\begin{table}[!h]
\centering
\caption{Three-fold stratified group cross-validation performance for three-class (HC, UC, CD) and two-class (HC vs. IBD) problems.}
\label{tab:performance}
\begin{tabular}{lcc}
\hline
\textbf{Performance Metric} & \textbf{Three classes} & \textbf{Two classes} \\
\textbf{(Mean $\pm$ Std. Dev.)} & \textbf{(HC, UC, CD)} & \textbf{(HC, IBD)} \\
\hline
Accuracy   & 0.774 $\pm$ 0.161 & 0.916 $\pm$ 0.118 \\
Precision  & 0.743 $\pm$ 0.220 & 0.927 $\pm$ 0.104 \\
Recall     & 0.741 $\pm$ 0.183 & 0.908 $\pm$ 0.129 \\
F1 Score   & 0.712 $\pm$ 0.209 & 0.912 $\pm$ 0.125 \\
\hline
\end{tabular}
\end{table}

\subsection{Model explainability analysis}
\subsubsection{Causal discovery}

\begin{figure}[!h]
    \centering
    \begin{subfigure}[b]{0.48\textwidth}
        \includegraphics[height=0.22\textheight, keepaspectratio]{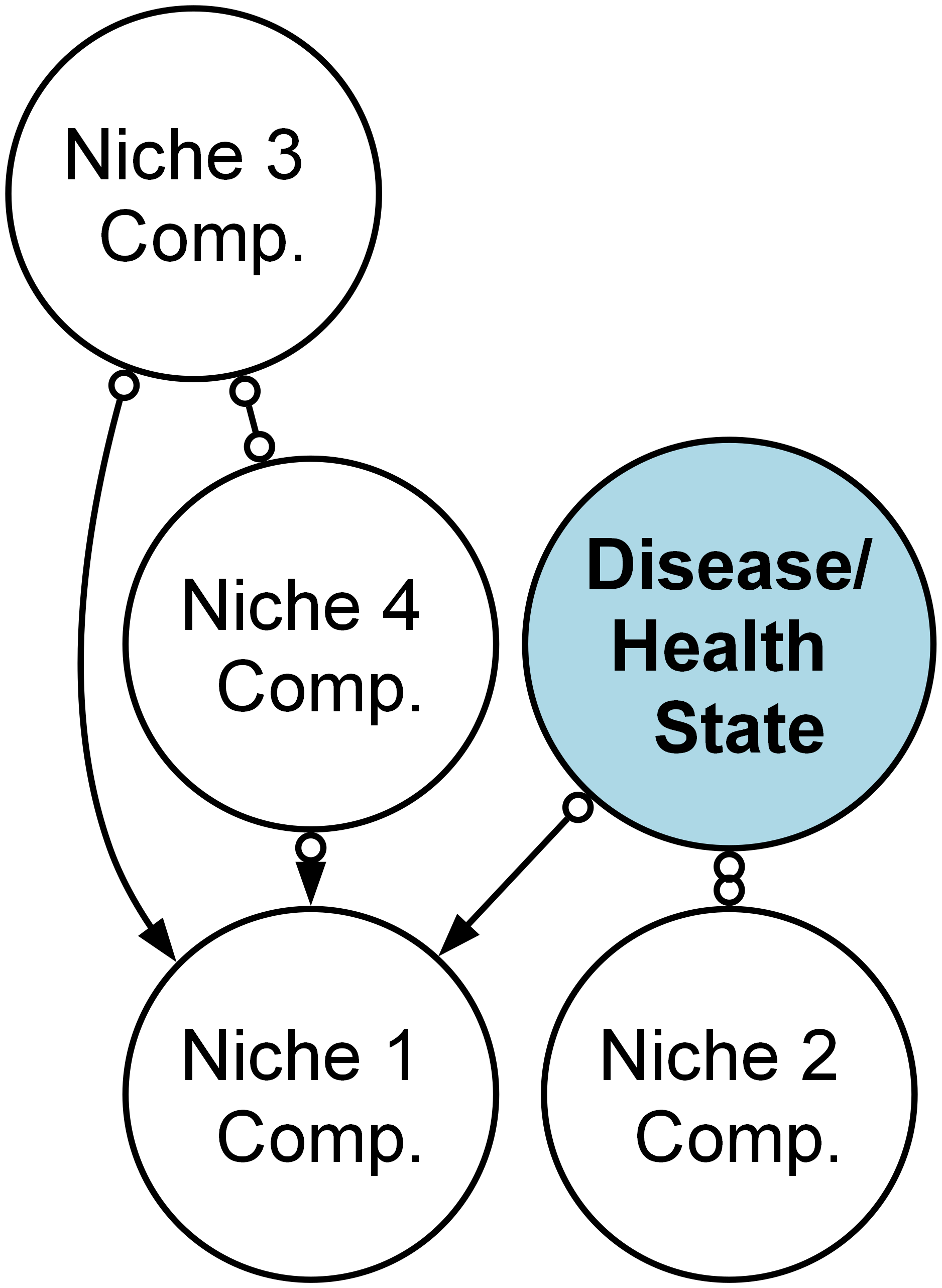}
        \caption{} 
        \label{fig:nichecomp}
    \end{subfigure}
    \hfill
    \begin{subfigure}[b]{0.48\textwidth}
        \includegraphics[height=0.24\textheight, keepaspectratio]{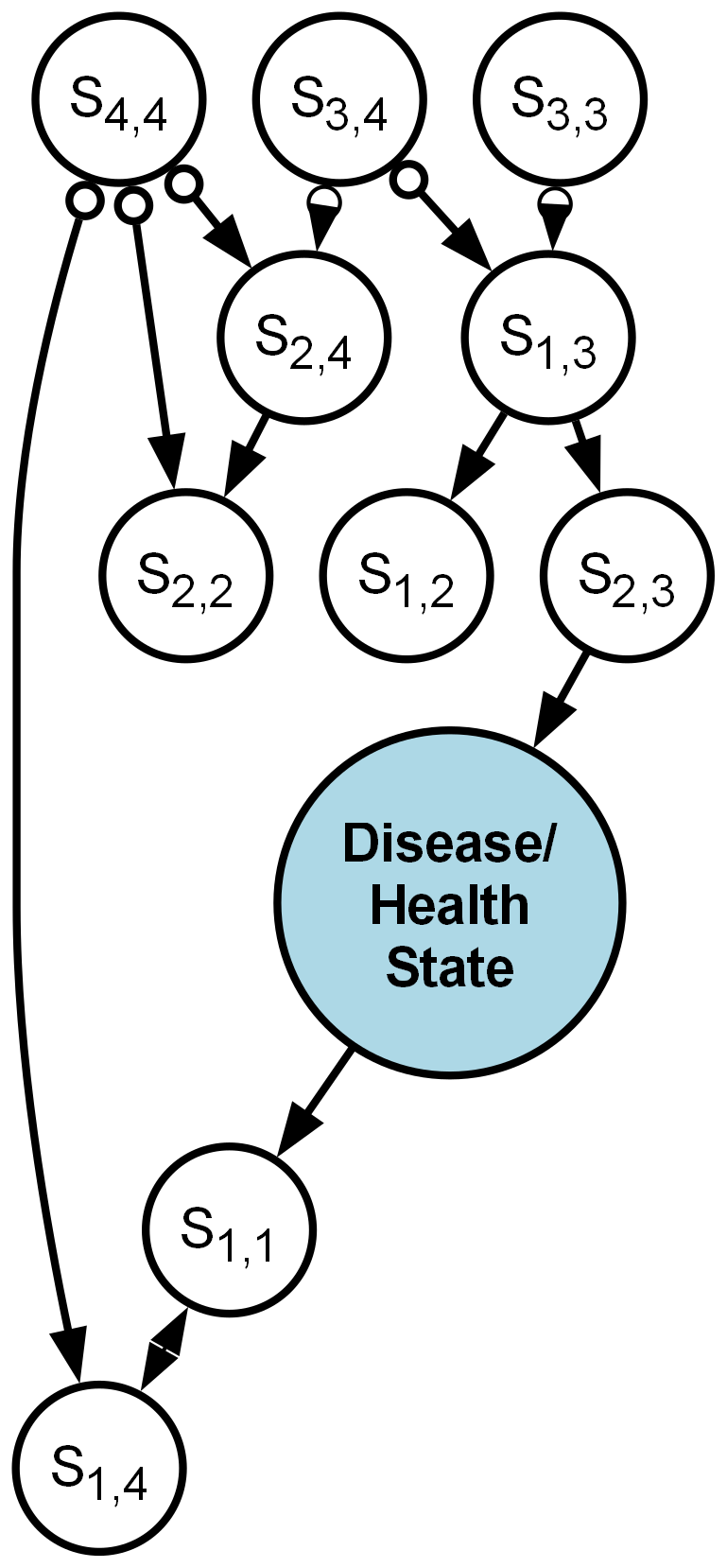}
        \caption{} 
        \label{fig:enrichment}
    \end{subfigure}

    \begin{subfigure}[b]{\textwidth}
        \includegraphics[width=0.9\textwidth]{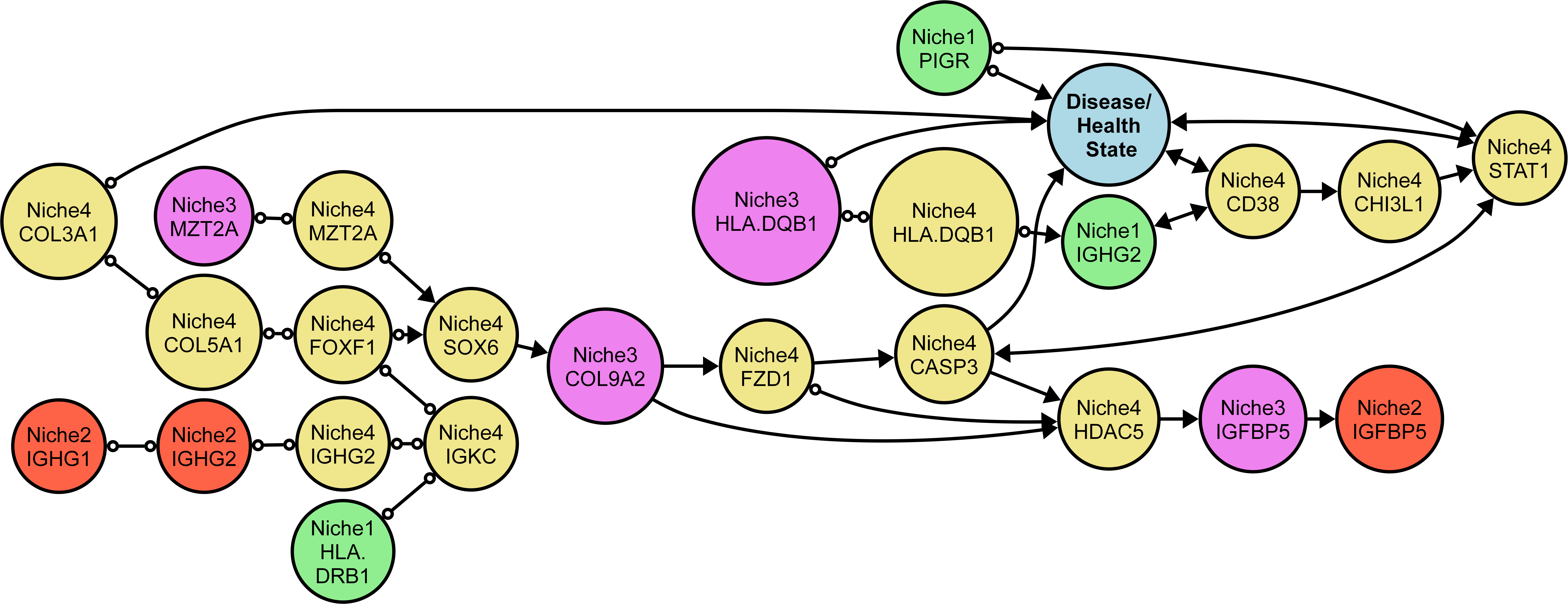}
        \caption{} 
        \label{fig:nichegene}
    \end{subfigure}

    \caption{Causal graphs depicting relationships between features and condition (Disease/Health State): (a) Niche composition, (b) Enrichment score, and (c) Niche-gene features. An open circle in an edge represents the potential presence of a latent confounder.}
    \label{fig:all}
\end{figure}

The causal graph in Figure~\ref{fig:nichecomp} shows that condition (Disease/Health State) is linked to Niche 1 and Niche 2 compositions, with edges suggesting either influence of disease on these niches or shared unmeasured causes. Niches 3 and 4 appear upstream of Niche 1, indicating that changes in these niches may precede or regulate downstream alterations. The relationship between Niches 3 and 4 remains unresolved, suggesting potential feedback or confounding.

In Figure~\ref{fig:enrichment}, two spatial interaction features have direct connections to disease status, highlighting specific niche–niche relationships as strong indicators of health versus disease. Other edges between enrichment features suggest a structured network of spatial dependencies, with some interactions influencing others and a few connections possibly driven by latent variables.

Finally, in Figure~\ref{fig:nichegene}, multiple niche–gene features show direct or confounded links to disease, as well. Disease status has bidirected connections with \texttt{Niche4\_CD38} and \texttt{Niche4\_STAT1}, as well as partially oriented edges with \texttt{Niche3\_HLA.DQB1}, \texttt{Niche4\_COL3A1}, and \texttt{Niche1\_PIGR}, indicating closely linked associations. Disease status also appears to be directly caused by \texttt{Niche4\_CASP3}. Moreover, within Niche 4, several features form a densely connected subnetwork. The overall structure indicates a complex network of interrelated niche–gene features involving multiple niches, genes and the condition (Disease/Health State).

\subsubsection{Statistical tests}

Table~\ref{tab:nmf_comparison_obs_clean} shows a significantly higher proportion of Niche 1 cells in HC than in UC (\(p \approx 0.0003\)) and CD (\(p < 0.0001\)); a significantly higher proportion of Niche 2 cells in UC than in HC (\(p < 0.0001\)) and CD (\(p \approx 0.0219\)); a significantly lower proportion of Niche 2 cells in HC than in UC (\(p < 0.0001\))  and CD (\(p < 0.0001\)); a significantly higher proportion of Niche 3 cells in UC than in HC (\(p < 0.0001\)) and CD (\(p < 0.0001\)); and a significantly higher proportion of Niche 4 cells in CD than in HC (\(p < 0.0001\)) and UC (\(p < 0.0001\)).

Table~\ref{tab:pairwise_interactions_means_obs} shows a significantly lower $\mathrm{S}_{1,1}$ in UC than in HC (\(p = 0.032\)) and CD (\(p = 0.012\)); a significantly higher  $\mathrm{S}_{1,2}$ in HC than in UC (\(p < 0.001\))  and CD (\(p = 0.014\)); and a significantly higher $\mathrm{S}_{2,1}$ in HC than in UC (\(p = 0.001\))  and CD (\(p = 0.014\)).   

Only statistically significant results are presented in both tables, with non-significant findings omitted for clarity.

\begin{table}[!h]
\centering
\caption{Pairwise comparisons of niche compositions across condition groups.}
\label{tab:nmf_comparison_obs_clean}
\resizebox{\textwidth}{!}{%
\begin{tabular}{|c|c|c|c|c|c|>{\centering\arraybackslash}p{4.7cm}|}
\hline
\textbf{NMF Factor} & 
\textbf{Group 1} & 
\textbf{Group 2} & 
\makecell[c]{\textbf{Difference}\\\textbf{Between}\\\textbf{Means}} & 
\textbf{Direction*} & 
\textbf{\textit{p}-value} & 
\textbf{Observation} \\
\hline

\multirow{2}{*}{1} 
 & HC & UC &  7.2503  & Up   & 0.0003  & \multirow{2}{*}{\centering HC-associated increase} \\[-0.2em]
 & CD & HC & -10.2075 & Down & <0.0001 & \\

\hline

\multirow{3}{*}{2} 
 & HC & UC & -4.4167  & Down & <0.0001 & \multirow{3}{*}{\centering\makecell{HC-associated decrease;\\UC-associated increase}} \\[-0.2em]
 & CD & HC &  2.4362  & Up   & <0.0001 & \\[-0.2em]
 & CD & UC & -1.9805  & Down & 0.0219  & \\

\hline

\multirow{2}{*}{3} 
 & HC & UC & -5.9405  & Down & <0.0001 & \multirow{2}{*}{\centering UC-associated increase} \\[-0.2em]
 & CD & UC & -7.3289  & Down & <0.0001 & \\

\hline

\multirow{2}{*}{4} 
 & CD & HC &  9.1598  & Up   & <0.0001 & \multirow{2}{*}{\centering CD-associated increase} \\[-0.2em]
 & CD & UC & 12.2667  & Up   & <0.0001 & \\

\hline
\end{tabular}
}

\parbox{\textwidth}{\scriptsize \textbf{*Note:} "Up" means Group 1 > Group 2.}
\end{table}

\begin{table}[!h]
\centering
\caption{Pairwise comparisons of niche interactions across condition groups.}
\label{tab:pairwise_interactions_means_obs}
\resizebox{\textwidth}{!}{%
\begin{tabular}{|c|c|c|c|c|c|c|c|}
\hline
\textbf{Interaction} & \textbf{Group 1} & \textbf{Group 2} & \textbf{Direction*} & 
\textbf{Group 1 Mean} & \textbf{Group 2 Mean} & \textbf{Adj. \textit{p}-value} & \textbf{Observation} \\
\hline

\multirow{2}{*}{1 vs 1}
& HC & UC & Up   &  0.165 & -2.627 & 0.032   & \multirow{2}{*}{UC-associated decrease} \\[-0.2em]
& UC & CD & Down & -2.627 & -0.039 & 0.012   & \\
\hline

\multirow{2}{*}{1 vs 2}
& HC & UC & Up   & -1.308 & -3.652 & <0.001 & \multirow{2}{*}{HC-associated increase} \\[-0.2em]
& HC & CD & Up   & -1.308 & -2.650 & 0.014  & \\
\hline

1 vs 4
& HC & UC & Up   & -1.612 & -4.851 & 0.002  & None \\
\hline

\multirow{2}{*}{2 vs 1}
& HC & UC & Up   & -1.693 & -3.673 & 0.001  & \multirow{2}{*}{HC-associated increase} \\[-0.2em]
& HC & CD & Up   & -1.693 & -2.917 & 0.014  & \\
\hline

4 vs 1
& HC & UC & Up   & -1.736 & -4.823 & 0.002  & None \\
\hline
\end{tabular}
}

\parbox{\textwidth}{\scriptsize \textbf{*Note:} "Up" means Group 1 > Group 2.}
\end{table}

\subsubsection{Feature importance analysis}

Among the 44 input features, only the top 20 are shown in Figure~\ref{fig:feature_importance_plots} (Appendix). In the three-class case, niche–gene features dominated, led by \texttt{niche\_3\_gene\_MZT2A}, followed by several Niche 4 genes (\texttt{COL5A1}, \texttt{MZT2A}, \texttt{FZD1}, \texttt{CASP3}). Other high-ranking features included \texttt{niche\_3\_gene\_HLA-DQB1}, \texttt{niche\_1\_gene\_PIGR}, and \texttt{niche\_4\_gene\_COL3A1}, with few enrichment or composition variables (\texttt{comp\_niche\_1}, \texttt{comp\_niche\_4}) appearing. For the two-class task (HC vs. IBD), enrichment features dominated, especially \texttt{enrichment\_2-2} ($\mathrm{S}_{2,2}$), \texttt{enrichment\_3-1} ($\mathrm{S}_{3,1}$), and \texttt{enrichment\_2-3} ($\mathrm{S}_{2,3}$), while only a few niche–gene features ranked highly. Thus, separating UC from CD relied more on gene-level signals, whereas distinguishing HC from IBD depended on spatial interaction patterns. 

However, PI can be misleading under correlated or redundant features, especially for highly nonlinear learners (e.g., MLPs). In such cases, permuting a single correlated feature may over- or under-estimate its contribution due to shared information and interaction terms. We therefore interpret PI qualitatively and report it with its variability across permutations, noting this caveat.

\section{Discussion}

We sought to develop a MLP classifier capable of classifying IBD subtypes by engineering a multi-layered feature set from ST data. The model achieved $\sim77\%$ accuracy in the more challenging three-class problem, with perfect classification of HC and lower accuracy between UC and CD. This is expected since distinguishing between UC and CD is challenging. A central finding of our work, revealed by PI analysis, is the differential utility of our engineered features. PI analysis revealed that this task depended primarily on niche-gene features, highlighting subtype-specific molecular signals within these microenvironments. In comparison, the simpler two-class problem achieved $\sim92\%$ accuracy, which was derived mainly from niche neighborhood enrichment scores, indicating that spatial cellular organization is a strong marker of general intestinal inflammation.

Our computational framework moves beyond a \textit{black box} prediction. Its explainable nature, incorporating causal discovery and feature importance analyses, allows for direct biological interpretation of the features driving IBD classification. The reliance of our model on neighborhood enrichment to identify general inflammation provides quantitative evidence that the disruption of spatial cytoarchitecture is a fundamental hallmark of the disease. This work builds upon previous studies that have provided descriptive characterizations of the IBD spatial landscape by transforming these insights into a robust predictive model. The identification of distinct niche-gene features, such as \textit{CASP3} in Niche 4 cells and \textit{HLA-DQB1} in Niches 3 and 4 cells for UC, and collagen-associated genes like \textit{COL5A1} in Niche 4 cells for CD, offers new testable hypotheses about the specific molecular pathways that define these subtypes.

The primary strength of our methodology lies in its novel, three-tiered feature engineering strategy, which captures niche composition, spatial enrichment, and localized gene expression to provide a holistic view of the tissue state. The use of NMF to define cellular niches provided an unbiased, data-driven approach to deconstructing complex spatial data into functional units. However, the small cohort of nine samples (three per condition) limits generalizability. The analysis was also constrained by the 980-gene panel; a whole-transcriptome approach could reveal more biomarkers. In addition, we recognize the risk of overfitting from extensive hyperparameter tuning on small datasets~\cite{crossval}. To mitigate this, we employed subject-level group-stratified cross-validation, ensuring that FOVs from the same subject never appeared in both training and validation folds. We also used \(L^2\)  regularization and a compact feature space to constrain complexity. Nonetheless, given the limited number of subjects, the reported accuracies should be interpreted as cross-validated estimates within a proof-of-concept framework, pending external validation on larger, independent, and multi-center cohorts.

This framework has significant clinical and research implications. It offers a potential path toward a more objective, data-driven diagnostic tool to help resolve the ambiguity between UC and CD. The specific spatial patterns and molecular markers identified could form the basis of novel diagnostic assays. For research, our findings provide deeper insights into the distinct biological mechanisms driving these diseases. The causal graphs suggest testable hypotheses about how changes in niche composition and interactions drive pathology. However, these graphs can only model linear interactions. Future studies could explore how these niche features evolve with therapy, potentially serving as predictive biomarkers for treatment outcomes, and should include functional investigations into the role of top-ranking genes in IBD pathogenesis.

\section{Conclusion}

By deconstructing colonic tissue into four cellular niches, we created a feature set capturing niche composition, spatial organization, and localized gene expression. Our model successfully addressed the three-class problem, revealing a fundamental principle of IBD pathology: subtype-specific molecular signals within these cellular niches are key to distinguishing UC from CD. In contrast, the simpler two-class task of separating IBD from HC highlighted that disruption in spatial tissue architecture is a primary indicator of general inflammation. Although we did not evaluate the model on the direct two-class problem to distinguish subtypes (UC versus CD), this represents an important next step. This study serves as a proof-of-concept, demonstrating how complex spatial data can be transformed into a robust, explainable, and predictive diagnostic framework, offering a pathway toward more precise diagnostics and targeted therapeutic strategies.

\appendix
\section{Appendix}

\begin{figure}[htbp]
    \centering
    \includegraphics[width=\linewidth]{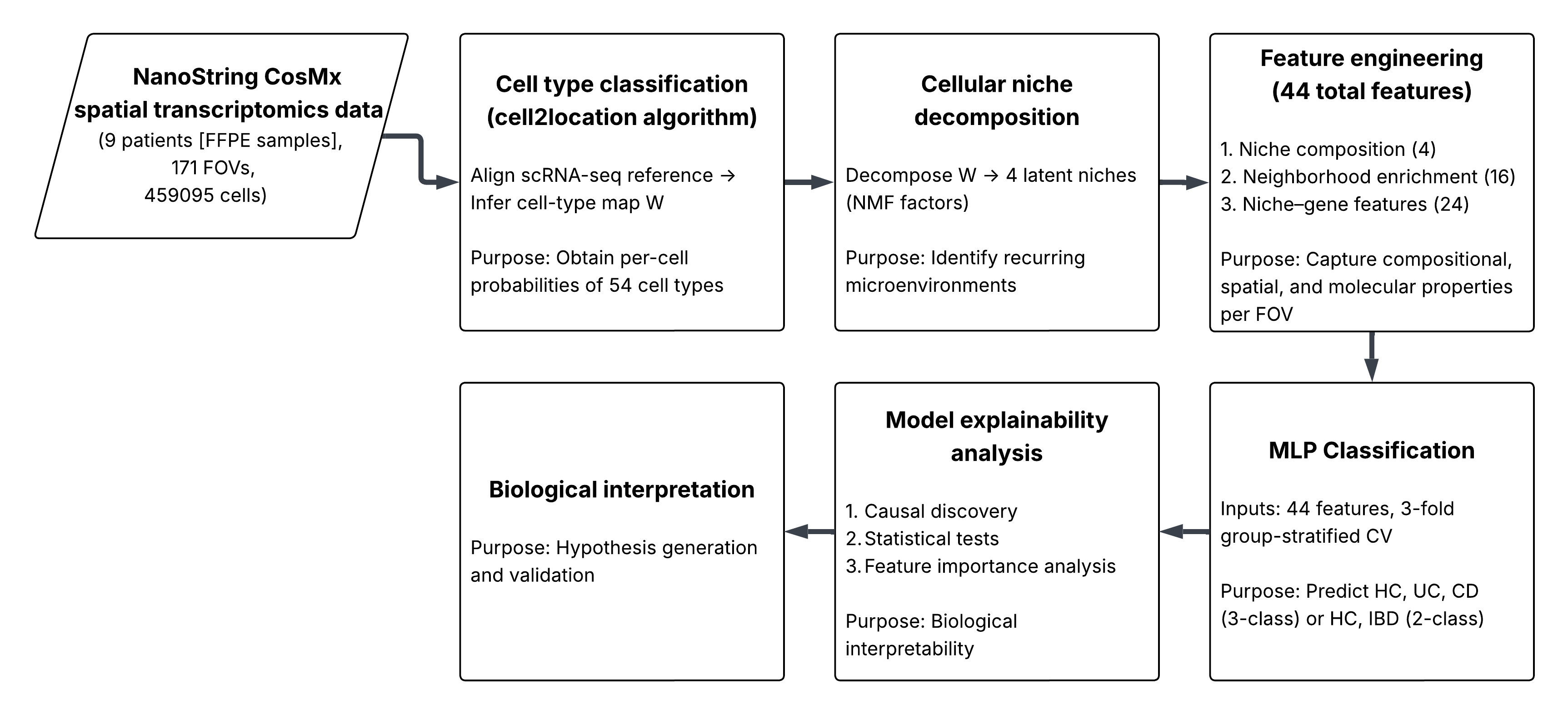}
    \caption{Overview of the computational pipeline. Starting from 
    NanoString CosMx spatial transcriptomics data (9 subjects, 171 FOVs, 
    459{,}095 cells), cell types were inferred using the 
    \texttt{cell2location} algorithm, and NMF to identify four recurrent cellular niches. From these niches, 44 features were engineered describing composition, spatial organization, and gene expression. A MLP classifier 
    trained using group-stratified cross-validation was then used to predict 
    disease states (HC vs. UC vs. CD or HC vs. IBD). Model explainability analyses (causal discovery, statistical tests, and permutation importance) enabled biological interpretation of the compositional, spatial and molecular determinants of IBD.}
    \label{fig:IBD_flowchart}
\end{figure}

\begin{figure}[H]
\centering
\begin{subfigure}[b]{0.48\textwidth}
    \centering
    \includegraphics[width=\textwidth]{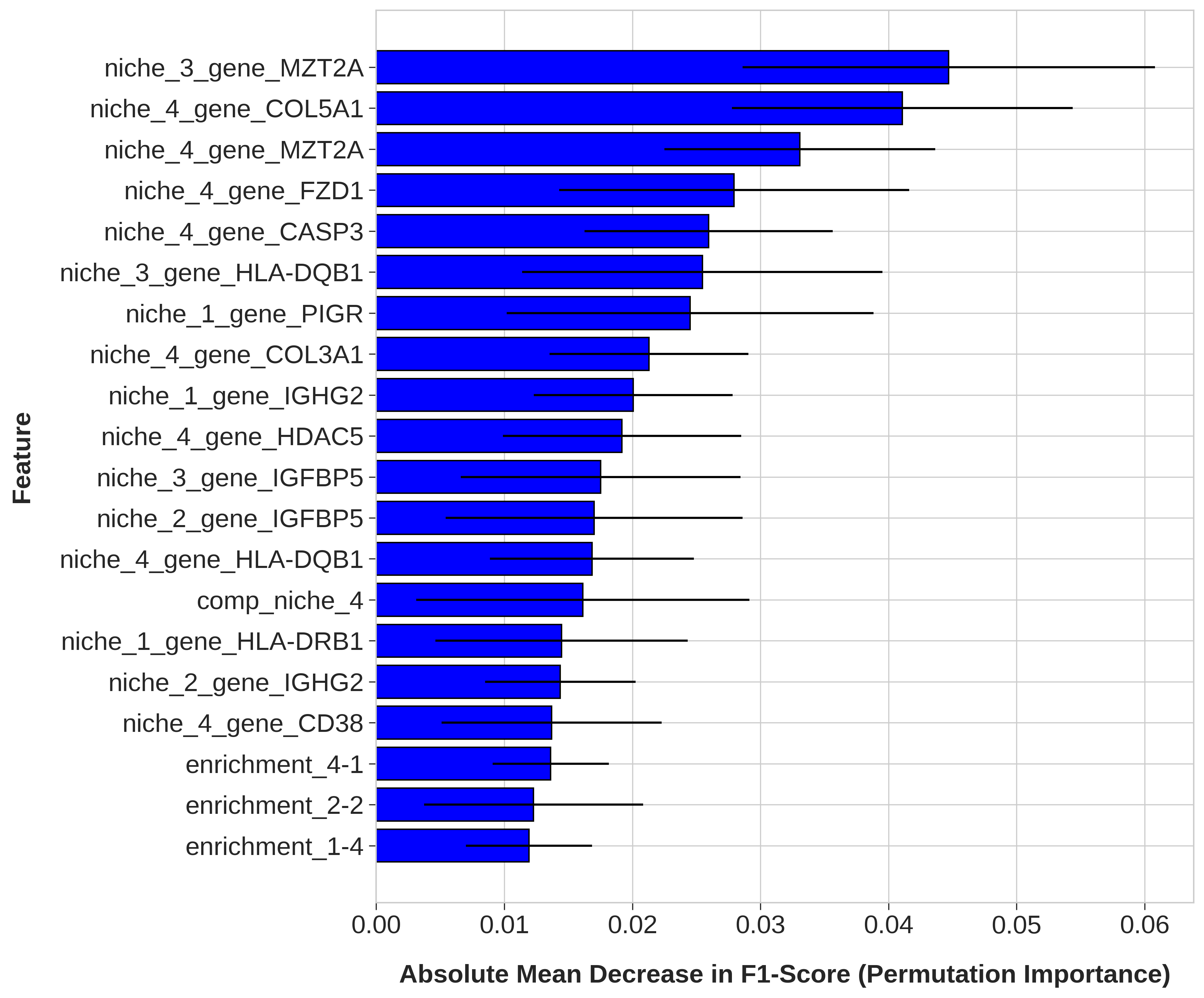}
    \caption{}
    \label{fig:perm_importance_3class}
\end{subfigure}
\hfill
\begin{subfigure}[b]{0.48\textwidth}
    \centering
    \includegraphics[width=\textwidth]{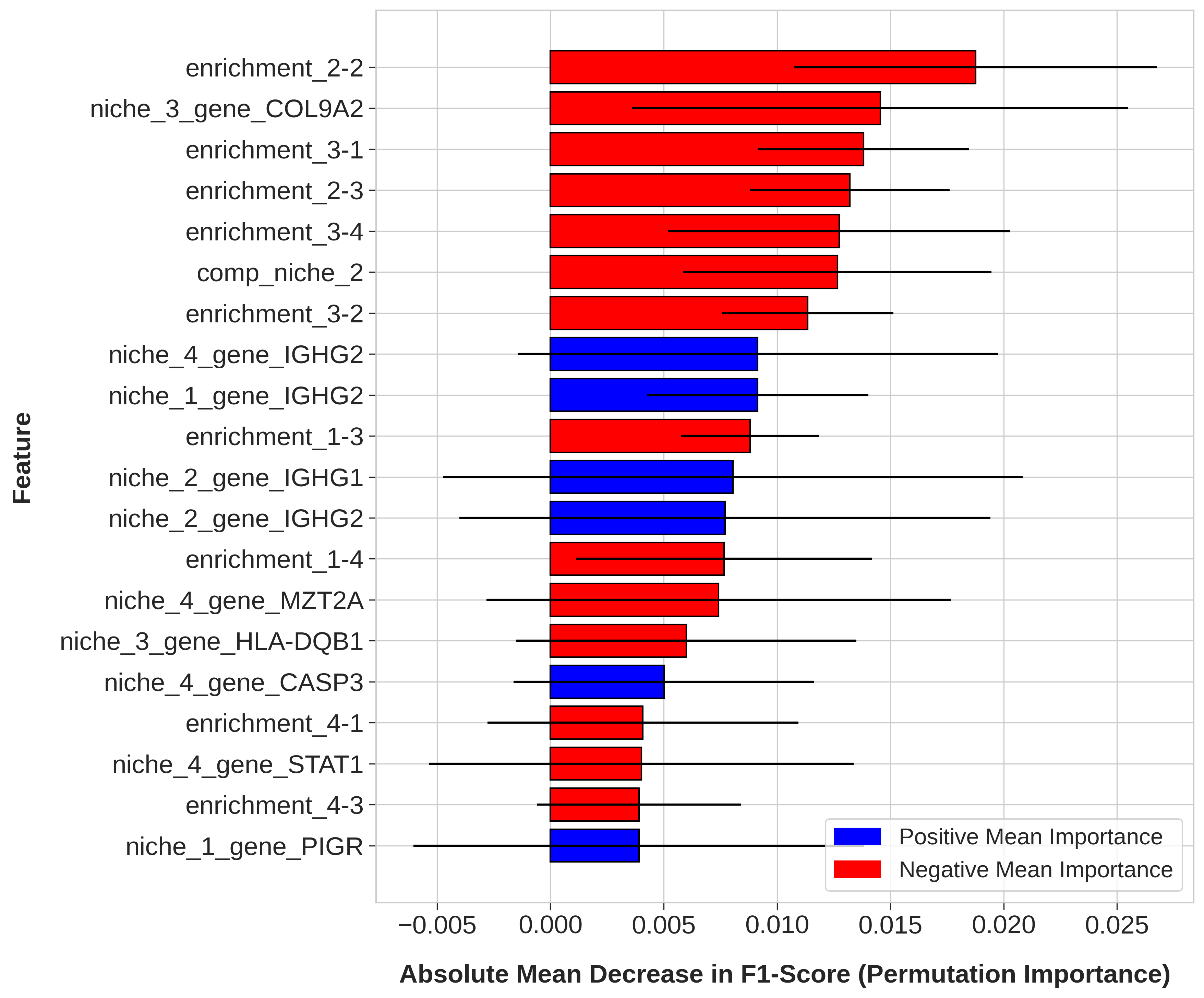}
    \caption{}
    \label{fig:perm_importance_binary}
\end{subfigure}
\caption{Top 20 features ranked by PI for the (a) three-class and (b) two-class problems.}
\label{fig:feature_importance_plots}
\end{figure}

\begin{credits}
\subsubsection{\ackname} 
This work was supported by the National Institutes of Health (NIH) grants R01HL127349, R01HL159805, R01HL178032. ChatGPT (GPT-5) was used solely to optimize phrasing and reduce wordiness in order to meet the page limits set by the conference. All ideas, analyses, and conclusions presented in this manuscript are entirely our own.

\subsubsection{\discintname}
The authors have no competing interests to declare that are relevant to the content of this article.
\end{credits}
%
%
%
%

\end{document}